\documentclass[universe,article,accept,moreauthors,pdftex]{Definitions/mdpi} 

\firstpage{1} 
\pubvolume{1}
\issuenum{1}
\articlenumber{0}
\pubyear{2022}
\copyrightyear{2022}
\externaleditor{{Academic Editors: Rosario Gianluca Pizzone and Sara Palmerini} 
} 
\datereceived{10 January 2022} 
\dateaccepted{07 February 2022} 
\datepublished{} 
\hreflink{https://doi.org/} 

\def\msun{M$_\odot$}
\def\iso#1{$^{#1}$}

\newcommand{\aap}{{\it Astron. Astrophys.}}
\newcommand{\mnras}{{\it Mon. Not. R. Astron. Soc.}}

\newcommand{\araa}{{\it Ann. Rev. Astron. Astrophys.}}

\newcommand{\pasa}{{\it Publ. Astron. Soc. Austr.}}
\newcommand{\gca}{{\it Geochim. Cosmochim. Acta}}

\newcommand{\ssr}{{\it Space Sci. Rev.}}
\usepackage{ textcomp }

\Title{The RADIOSTAR Project}

\TitleCitation{The RADIOSTAR Project}

\Author{{Maria Lugaro} 
 $^{1,2,3,*}$\orcidA{}, {Benoit C\^ot\'e $^{1,4,5,\ddagger}$\orcidB{},} 
 Marco Pignatari~$^{1,6,4,\ddagger}$, {Andr\'es Yag\"ue L\'opez} 
 $^{1,\dagger,\ddagger}$\orcidD{}, Hannah Brinkman $^{1,7}$, Borb\'ala Cseh $^{1}$, 
Jacqueline Den Hartogh $^{1}$\orcidG{}, {Carolyn Louise Doherty} $^{1}$, 
{Amanda Irene Karakas} $^{3,8}$, Chiaki Kobayashi $^{9}$\orcidK{}, Thomas Lawson $^{1,6}$\orcidT{}, 
{M\'aria Pet\H{o}} $^{1}$, {Benj\'amin So\'os} $^{1,2}$, 
Thomas Trueman $^{1,6}$ and {Blanka Vil\'agos} $^{1,2}$}

\AuthorNames{Maria Lugaro, Benoit C\^ot\'e, Marco Pignatari, Andr\'es Yag\"ue L\'opez, Hannah Brinkman, Borb\'ala Cseh, Jacqueline Den Hartogh, Carolyn Doherty, Amanda I. Karakas, Chiaki Kobayashi, et al.}

\AuthorCitation{Lugaro, M.; C\^ot\'e, B.; Pignatari, M.; Yag\"ue L\'opez, A.; Brinkman, H.; Cseh, B.; Den Hartogh, J.; Doherty, C.L.; Karakas, A.I.; Kobayashi, C.; et al.}

\address{%
$^{1}$ \quad Konkoly Observatory, Research Centre for Astronomy and Earth Sciences, E\"otv\"os Lor\'and Research Network (ELKH), Konkoly Thege Mikl\'{o}s \'{u}t 15-17, {H-1121} 
 Budapest, Hungary; benoit.cote@csfk.org (B.C.); marco.pignatari@csfk.org (M.P.); andres.yague@csfk.org (A.Y.L.); hannah.brinkman@csfk.org (H.B.); cseh.borbala@csfk.org (B.C.); jacqueline.den.hartogh@csfk.org (J.D.H.); carolyn.louise.doherty@gmail.com~(C.D.); T.Lawson-2012@hull.ac.uk (T.L.); maria.k.peto@gmail.com (M.P.); soos.benjamin@csfk.org (B.S.); T.Trueman-2018@hull.ac.uk (T.T.); vilagos.blanka@csfk.org (B.V.)\\
$^{2}$ \quad {ELTE Institute of Physics,} 
 E\"{o}tv\"{o}s Lor\'and University, P\'azm\'any P\'eter S\'et\'any 1/A, {1117} Budapest, Hungary\\
$^{3}$ \quad School of Physics and Astronomy, Monash University, VIC 3800, Australia; amanda.karakas@monash.edu\\
$^{4}$ \quad Joint Institute for Nuclear Astrophysics---Center for the Evolution of the Elements  (JINA-CEE), University of Notre Dame, South Bend 46556, IN,{USA}\\
$^{5}$~~~~Department of Physics and Astronomy, University of Victoria, Victoria, BC V8P 5C2, Canada

$^{6}$ \quad \textls[-15]{E.~A.~Milne Centre for Astrophysics, Department of Physics and Mathematics, University of Hull, {Kingston upon Hull HU6 7RX,} 
 UK}\\
$^{7}$ \quad Graduate School of Physics, University of Szeged, Dom t\'er 9, 6720 Szeged,  Hungary\\
$^{8}$ \quad ARC Centre of Excellence for All Sky Astrophysics in 3 Dimensions {(ASTRO 3D), Australia} 
\\
$^{9}$ \quad Centre for Astrophysics Research, {University of Hertfordshire, College Lane,} 
 Hatfield AL10 9AB, UK; c.kobayashi@herts.ac.uk
}

\corres{Correspondence: maria.lugaro@csfk.org}

\firstnote{Current address: Computer, Computational and Statistical Sciences (CCS) Division, Center for Theoretical Astrophysics, Los Alamos National Laboratory, Los Alamos, NM 87545, USA.} 
\secondnote{NuGrid Collaboration, http://nugridstars.org}

\abstract{ 
Radioactive nuclei are the key to understanding the circumstances of the birth of our Sun because meteoritic analysis has proven that many of them were present at that time. Their origin, however, has been so far elusive. The ERC-CoG-2016 RADIOSTAR project is dedicated to investigating the production of radioactive nuclei by nuclear reactions inside stars, their evolution in the Milky Way Galaxy, and their presence in molecular clouds. So far, we have discovered that: (i) radioactive nuclei produced by {$slow$}
 (\iso{107}Pd and \iso{182}Hf) and $rapid$ (\iso{129}I and \iso{247}Cm) neutron captures originated from 
stellar sources {---asymptotic giant branch (AGB) stars and compact binary mergers, 
respectively---} {within}
 the galactic environment that predated the formation of the molecular cloud where the Sun was born; (ii) the time that elapsed from the birth of the cloud to the birth of the Sun was of the order of 10$^7$ years, and (iii) the abundances of the very short-lived nuclei \iso{26}Al, \iso{36}Cl, and \iso{41}Ca can be explained by massive star winds in single or binary systems, if these winds directly polluted the early Solar System. Our current and future work, as required to finalise the picture of the origin of radioactive nuclei in the Solar System, involves studying the possible origin of radioactive nuclei in the early Solar System from core-collapse supernovae, investigating the production of \iso{107}Pd in massive star winds, modelling the transport and mixing of radioactive nuclei in the galactic and molecular cloud medium, and calculating the galactic chemical evolution of \iso{53}Mn and \iso{60}Fe and of the $p$-process isotopes \iso{92}Nb and \iso{146}Sm.}


\begin{document}

\section{Introduction}
\label{sec:intro}

High-precision analysis of meteoric rocks and inclusions has allowed us to discover an intriguing
property of our Solar System: at its birth, 4.6 billion years ago, it was rich in radioactive nuclei. Most of them were short-lived, with half-lives (T$_{1/2}$) less than 100~million years (Myr), and became extinct within the first few hundreds Myr of the Solar System evolution. We can still infer their initial abundances by analysing meteorites for excesses in the abundances of the daughter nuclei into which they decay. For example, if a mineral has a high \iso{27}Al/\iso{24}Mg ratio and a \iso{26}Mg/\iso{24}Mg ratio higher than solar, then the \iso{26}Mg excess must have been originally incorporated in the mineral as \iso{26}Al, which behaves chemically as \iso{27}Al but is radioactive and has a half-life of 0.7 Myr. From the correlation between the \iso{27}Al/\iso{24}Mg and the \iso{26}Mg/\iso{24}Mg ratios, a relatively high  \iso{26}Al/\iso{27}Al$\sim$$5 \times 10^{-5}$ is derived for the time when the Sun was born \cite{lee77,jacobsen08}. Therefore, the decay of \iso{26}Al was the major source of heating in the planetesimals that accreted within the first few million years in the early Solar System (ESS). This heat led to melting and differentiation \cite{moskovitz11}, as well as water circulation and loss in those planetesimals that were ice rich \cite{ciesla15,lichtenberg19}.

On top of \iso{26}Al, there is convincing evidence of the presence in the ESS of 
another ten short-lived radioactive (SLR) nuclei produced in stars and explosive environments: \iso{36}Cl (T$_{1/2}$=0.3 Myr), \iso{53}Mn (T$_{1/2}$=3.7 Myr), \iso{60}Fe (T$_{1/2}$=2.6 Myr), \iso{92}Nb (T$_{1/2}$=34.7 Myr), \iso{107}Pd (T$_{1/2}$=6.7 Myr), \iso{129}I (T$_{1/2}$=15.7 Myr), \iso{146}Sm (T$_{1/2}$=68 or 103 Myr, debated), \iso{182}Hf (T$_{1/2}$=8.9 Myr), \iso{244}Pu (T$_{1/2}$=80 Myr), and \iso{247}Cm (T$_{1/2}$=15.6 Myr). These are listed in Table~2 of \cite{lugaro18rev}, together with their half-lives, the stable isotopes used as reference to derive their abundances, and the abundance ratios inferred for the ESS. {{For another six isotopes},}
including \iso{41}Ca (T$_{1/2}$=0.1 Myr), weaker evidence or only upper limits exist {for their ESS abundances}. 
The physical properties of these SLR nuclei cover wide ranges: half-lives are from 0.1 to 100 Myr and masses from very light (\iso{26}Al) to extremely heavy (\iso{247}Cm). Production of these nuclei in stars is ascribed to a large variety of nuclear physics processes: from proton captures (e.g., \iso{26}Al), to neutron captures (e.g., \iso{129}I), disintegration of heavier nuclei (e.g., \iso{146}Sm), and nuclear statistical equilibrium (e.g., \iso{53}Mn). {These processes occur} in different stellar environments, from core-collapse and thermonuclear supernovae to giant stars. Moreover, the accuracy and precision of the SLR abundances in the ESS vary from a few percent (e.g., for \iso{26}Al and \iso{182}Hf) to values debated by orders of magnitudes. {For example,} the value of \iso{60}Fe/\iso{56}Fe is most likely to be 
low, $\sim$$10^{-8}$ \cite{tang15,trappitsch18}, but higher values up to $\sim 10^{-6}$ have been reported too \cite{mishra14}.

The origin of SLR nuclei in the ESS, and consequently their potential presence in extrasolar planetary systems, is still poorly understood and hotly debated. The different scenarios have implications on the circumstances and the sequence of events that led to the birth of the Sun \cite{adams10,pfalzner15}
and on heating and ionisation in star-forming regions \cite{adams14,lacki14}. Furthermore, the decay of \iso{26}Al, and potentially \iso{60}Fe, was an important source of heat inside the first planetesimals. {Therefore,} their presence or absence determines several important properties of extrasolar planetesimals, including water circulation and loss for those planetesimals that are ice rich. This, in turn, determines not only mineral diversity and modification of organic molecules, as the presence of water drives hydro-thermal processing, but also the amount of water delivered by these planetesimals to planets in the habitable zone. If \iso{26}Al is present, a high fraction of water is mobilised and removed from ice-rich planetesimals and mainly dry planets result in the habitable zone \cite{raymond07}. Without \iso{26}Al, instead, ice-rich planetesimals keep their initial ice content and deliver more water to planets in the habitable zone \cite{ciesla15,lichtenberg19}. A vast amount of literature (e.g., \cite{cameron77,wasserburg96,busso99,meyer00,wasserburg06,huss09,gaidos09,gounelle12,pan12,young14,lichtenberg16b,dwarkadas17,fujimoto18}) in the past 40~years has been devoted to try to solve the puzzle of radioactivity in the ESS, as reviewed by Lugaro, Ott, and Kereszturi \cite{lugaro18rev}. A more recent review on the production specifically of $^{26}$Al and $^{60}$Fe and their astrophysics relevance can be found in \cite{diehl:21}.

\begin{figure}[H]
\includegraphics[width=9.5 cm]{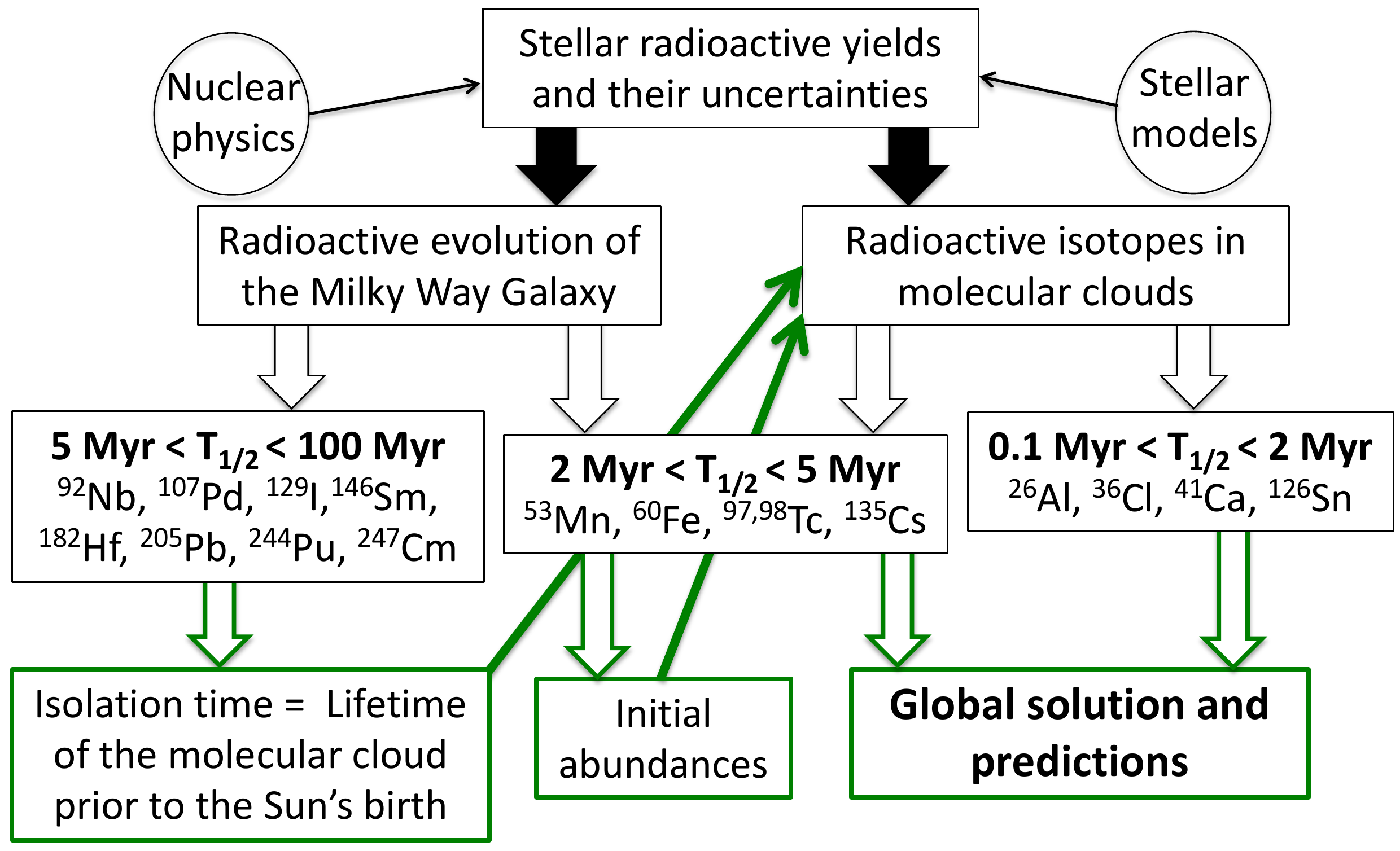}
\caption{Flowchart of the RADIOSTAR project. From top: stellar models and nuclear physics are used as input for calculating the stellar yields of SLR nuclei. {These,} in turn, serve as input for investigating the evolution of the SLR abundances in the Milky Way Galaxy and in molecular clouds. In first approximation, the half-life of each SLR nucleus (third row) determines if its abundance in the ESS was predominately contributed by stars in the Galaxy (over longer timescales), in molecular cloud (over shorter timescales), or in both. The abundances of the SLR nuclei with relatively long half lives (third row, left box) have a predominant galactic contribution. {They} can be used to determine the initial composition of the molecular cloud where the Sun was born and the timescale of such birth. This information can be fed (green arrows) into the analysis of the cloud evolution. \label{fig:scheme}}
\end{figure}

The {RADIOSTAR} 
 \endnote{``Radioactivities from Stars to Solar Systems'', \url{https://konkoly.hu/radiostar/ }{(accessed on 1 January 2022)} }project was funded by the European Research Council 
(ERC-CoG-2016) starting from September 2017. {Its objective is} to produce the first complete, self-consistent picture of the origin of SLR nuclei in the ESS from the stars that produced them. The two main goals of RADIOSTAR are (see Figure~\ref{fig:scheme}): (1) To derive the time that elapsed from the formation of the molecular cloud to the formation of the Sun (the ``isolation'' time). The abundances of those SLR nuclei that are sensitive to galactic timescales {can be used to this aim}. The isolation time also represents the timescale of the potential process of self-pollution by stellar ejecta within the molecular cloud itself. (2) To consider all the SLR nuclei and analyse {all the mechanisms of their stellar production in all the possible different stellar sources}. The final objectives are to find global self-consistent solutions for all the SLR nuclei, derive the uncertainty space where such solutions are valid, and provide testable predictions for the abundances of those SLR nuclei whose abundances in the ESS are not yet well known.

Here, we report the progress that has been made by the project since 2017, both in the stellar yield calculations (Section~\ref{sec:stars}) and in the modelling of the galactic chemical evolution of the SLR nuclei (Section~\ref{sec:galaxy}). In Section~\ref{sec:final}, we report on current and future work to be performed within the project, as it will run to its completion at the end of August 2023.

\section{Progress on Stellar Yields of Radioactive Isotopes}
\label{sec:stars}

Massive stars and their core-collapse supernovae (CCSNe) are one of the main sources of the SLR abundances measured in the ESS. Therefore, within the RADIOSTAR project, we dedicated a special effort to the study of the nucleosynthesis within these stellar objects. Present uncertainties affecting theoretical stellar simulations of massive stars are a major challenge for our analysis. In Lawson et al.~\cite{lawson:21}, we explored the production of SLR nuclei relevant for the ESS using 62 CCSN models with initial masses of 15, 20, and 25~M$_{\odot}$. {The models cover} a range of explosion energies between 3.4 $\times$ 10$^{50}$ and 1.8 $\times$ 10$^{52}$ ergs and different ejecta configurations \citep[][]{fryer18}. The same models were previously adopted to study the nucleosynthesis of radioactive isotopes relevant for future $\gamma$-ray astronomy observations~\citep[][]{andrews:20} and the nucleosynthesis of $^{60}$Fe \citep[][]{jones:19}. 
In \cite{lawson:21}, we highlighted the diversity and the wide range of production mechanisms seen for the SLR nuclei. {In some cases, such as $^{36}$Cl and $^{41}$Ca, isotopes} can be efficiently made in different
parts of the ejecta by different nucleosynthesis production paths and different nuclear reactions. In other cases, like $^{60}$Fe, the same nuclear reactions are triggered at different temperature and density conditions in the ejecta. We also showed that we cannot neglect the impact of nucleosynthesis prior to the CCSN. Isotopic abundances produced prior to the explosion, {for example,} for $^{26}$Al, $^{60}$Fe, and $^{182}$Hf, may also be ejected by the explosion, and the relevance of such a contribution with respect to explosive nucleosynthesis varies greatly between models. 

Our set of CCSN models \cite{lawson:21} allowed us to study the impact of both the explosion energy and the compact remnant mass, which varies between 1.5 M$_{\odot}$ and 
4.9 M$_{\odot}$ in this set of models \citep[][]{fryer18}. Isotopes such as $^{92}$Nb and $^{97}$Tc are significantly produced within the innermost ejecta, and therefore they are strongly affected by {the remnant mass, which is very uncertain}. In particular, for $^{92}$Nb, we obtain yields  varying by five orders of magnitude between different CCSN models of the same stellar progenitor.
An extensive comparison with other models available in the literature \citep[][]{rauscher02, sieverding:18, limongi18, curtis19} confirms or increases the scatter of SLR yields obtained in our study. For the yields of $^{60}$Fe and lighter SLR nuclei, with the exception of $^{41}$Ca, we find agreement within an order of magnitude across the {published} sets of models for different progenitors. Concerning heavier SLR nuclei, we obtain roughly an order of magnitude less $^{98}$Tc compared to {other models available in the literature}. The same applies to $^{126}$Sn, for the smaller stellar progenitors, 
the 15 M$_{\odot}$ and 20 M$_{\odot}$ stars. The source of such a discrepancy is still unclear, given the large parameter space covered in our study in terms of CCSN model setup. 
Our $^{92}$Nb median yield for the 15 M$_{\odot}$ models overproduce this SLR isotope by
at least three orders of magnitude compared to other mass models, due to its high production in models with small remnant mass and medium-high explosion energy. Another heavy SLR isotope, the proton-rich $^{146}$Sm, shows the largest variation between different stellar sets, of about two orders of magnitude. The other heavy SLR isotope yields considered in our study ($^{97}$Tc, $^{107}$Pd, $^{129}$I, $^{135}$Cs, $^{182}$Hf, and $^{205}$Pb) are consistent among the different masses within an order of magnitude. 

\begin{figure}[H]
\includegraphics[width=13.5 cm]{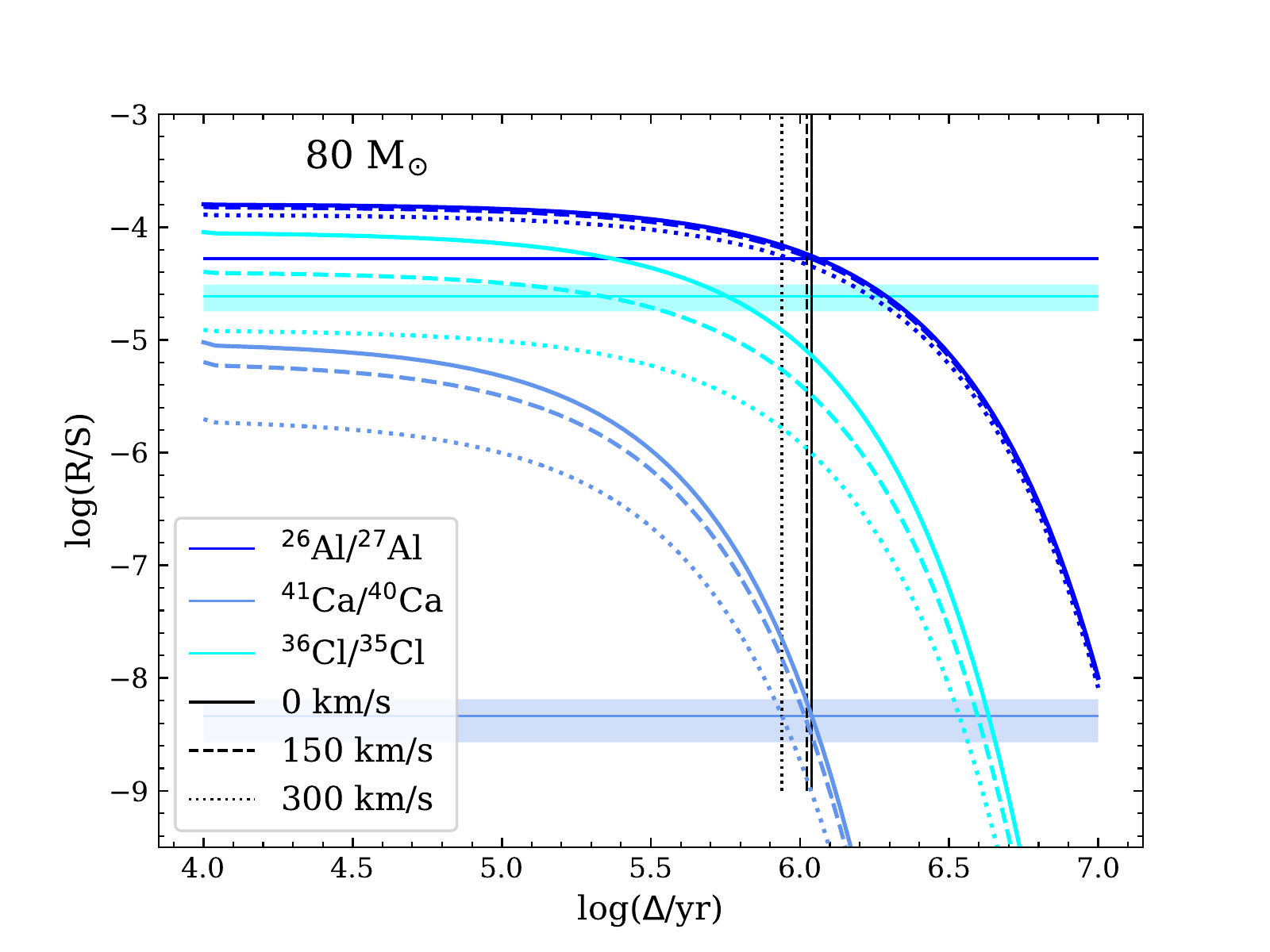}
\caption{{{Previously} 
published as panel e) of Figure 7 in \cite{brinkman21}, \copyright AAS, reproduced with
permission. Abundance ratios (R/S) for the SLR $^{26}$Al, $^{36}$Cl, and $^{41}$Ca over their stable 
reference isotope
($^{27}$Al, $^{35}$Cl, and $^{40}$Ca) for a dilution factor $f$ in the range $0.16-0.5\times 10^{-4}$. 
These values are calculated as $M_{\rm ESS}$($^{26}$Al)/$M_{\rm star}$($^{26}$Al), where $M_{\rm 
ESS}$($^{26}$Al) is the mass of $^{26}$Al observed in the ESS and $M_{\rm star}$($^{26}$Al) its stellar 
yield from the example stellar models of initial 80 \msun\ and different initial velocities as indicated. 
The horizontal bands represent the ESS ratios, with their respective errors. The vertical lines represent 
the delay time, i.e., the time interval needed to match the ESS $^{41}$Ca/$^{40}$Ca ratio, typically of the 
order of 1 Myr.}} \label{fig:winds}

\end{figure}

The SLR signatures are not the only isotopic anomalies detected in the ESS. In fact, several stable isotope variations of nucleosynthetic origin have been identified in meteoritic samples (see review by \cite{kleine20}), also possibly correlated to unstable isotopic abundances (e.g.,~\cite{larsen11}). The source of such variations is generally attributed to the presence {of} stardust grains {that predated the formation of the Solar System}. Since CCSNe are among the relevant producers of stardust grains (e.g., \cite{zinner14}), the same models used to study the production of SLR nuclei can be compared with stardust grains abundances and meteoritic anomalies. 
In den Hartogh et al. (2022, ApJ accepted \endnote{\url{https://arxiv.org/abs/2201.04692}(accessed on 1 January 2022)}), 
we used CCSN models to study stardust chromium-rich oxide grains recently measured by \cite{nittler:18}. {These grains are} possibly responsible for the observed $^{54}$Cr variations in materials formed in different regions of the proto-planetary solar disk (e.g., \cite{dauphas:10,qin:10}) and their correlation with \iso{26}Mg variations~\cite{larsen11}. By using three different sets of CCSN models, in den Hartogh et al., we showed that CCSN ejecta can reproduce the grain anomalous $^{54}$Cr/$^{52}$Cr, $^{53}$Cr/$^{52}$Cr {ratios observed in the chromium-rich grains}. {They can also match} the {measured} $^{50}$Cr/$^{52}$Cr (the C-ashes) or $^{50}$Ti/$^{48}$Ti~\endnote{The origin of the observed signal at mass 50 is still unknown, as it could come from Cr and Ti.} (the He ashes) ratios. Therefore, these grains, previously attributed to electron-capture supernovae~\cite{nittler:18,jones:19a}, could have been made by CCSNe. Finally, we found that in the CCSN models, \iso{54}Cr excesses are accompanied by excesses in the stable \iso{26}Mg, rather than in the SLR \iso{26}Al.
The stellar winds of massive Wolf--Rayet stars may also eject a significant amount of material in the {interstellar medium (ISM)} and are therefore also relevant for the production of SLR nuclei.   {References} 
~\cite{arnould97,arnould06} first provided a comprehensive analysis of the production of SLR nuclei in massive stars {and their ejection by} Wolf--Rayet winds, for both non-rotating and rotating stellar models. In Brinkman et al. \cite{brinkman19,brinkman21}, we presented an updated study of the production of the SLR nuclei in massive star winds, where we also considered the effect of binary interaction. Wind yields were calculated for stellar evolution models of solar metallicity in the mass range 10 M$_{\odot}$ $\leq$ M$\leq$ 80 M$_{\odot}$ for non-rotating single and binary stars (for $^{26}$Al \cite{brinkman19}) and for non-rotating and rotating single stars (for $^{26}$Al, $^{36}$Cl, and $^{41}$Ca~\cite{brinkman21}). The main results are that binary interactions increase the $^{26}$Al wind yields in
stars of masses 10-35 M$_{\odot}$, while for the more massive Wolf--Rayet stars the impact is small \cite{brinkman19}. Rotation does not affect significantly the $^{26}$Al yields but increases/decreases the $^{36}$Cl and $^{41}$Ca yields for stars in the lower/higher mass range. From about 45 M$_{\odot}$, the yields become again comparable between models with different initial conditions \cite{brinkman21}. {The full impact of the new models and their effect relative to the CCSN contribution in the Galaxy and specifically in star-forming regions can only be established by considering full models of population synthesis (see, e.g., \cite{voss09}).} Nevertheless, our two studies confirm that massive star winds are a realistic candidate for the origin of at least $^{26}$Al, $^{36}$Cl, and $^{41}$Ca in the ESS, as demonstrated by the example shown in Figure~\ref{fig:winds}.


\section{Progress on Galactic Chemical Evolution of Radioactive Isotopes}
\label{sec:galaxy}


The ESS abundances of SLR nuclei are typically given as an abundance ratio between an SLR nucleus and a stable or long-lived reference isotope. { The} SLR nuclei only encode the contribution of a few enrichment events, i.e., the last events that occurred prior to the formation of the Sun. {Therefore,} their abundances reflect the convolution between the stellar yields and the local star formation rate at the time of the ESS. Instead, because the stable reference isotopes do not decay, their abundances encode the contribution of all enrichment events that occurred prior to the formation of the Sun. {Therefore, they} reflect the complete chemical evolution history of the Galaxy, at least for the solar neighbourhood. To properly interpret the meteoritic abundances ratios, galactic chemical evolution (GCE) models and simulations must be employed, which incorporate nucleosynthesis and stellar evolution predictions into galactic frameworks that follow the star formation history of our Galaxy. The use of GCE not only allows us to predict the abundances of SLR nuclei in the ESS but to gain insights into their astrophysical origins and to quantify the isolation time.

Previous GCE studies developed analytical solutions for the abundances of SLR nuclei {in the ISM} by combining production yields ratios with parametrised star formation histories \cite{Clayton1984,clayton88,meyer00,huss09}. Building and extending upon these pioneering studies, we incorporated radioactivity in the GCE code OMEGA+ \cite{cote19PaperI}. This code is part of an open-source numerical pipeline \cite{cotepipeline,omegaplus} that offers several advantages over the use of analytical solutions. First, it has strong connections with the nuclear astrophysics community. {In fact}, it can directly incorporate nucleosynthetic yields predictions from a wide range of astrophysical sites such as CCSNe, low- and intermediate-mass stars, neutron star mergers and black hole--neutron star mergers, as well as different channels for supernovae Type Ia (SNeIa, see also~\cite{travaglio14}). Each set of yields can be mass and metallicity dependent. Second, our framework considers delay-time distributions, representing the probability of an enrichment event (e.g., an SNIa) to occur after a given time since the formation of its progenitor (e.g., a low-mass binary system). Third, the radioactivity in OMEGA+ is connected to a decay module that can accommodate radioactive isotopes with multiple decay channels. Finally, fourth, our GCE codes include important galaxy evolution processes that were not accounted for in previous analytical solutions, such as galactic outflows driven by the energy of supernova~explosions.

\begin{figure}[H]
\includegraphics[width=13.5 cm]{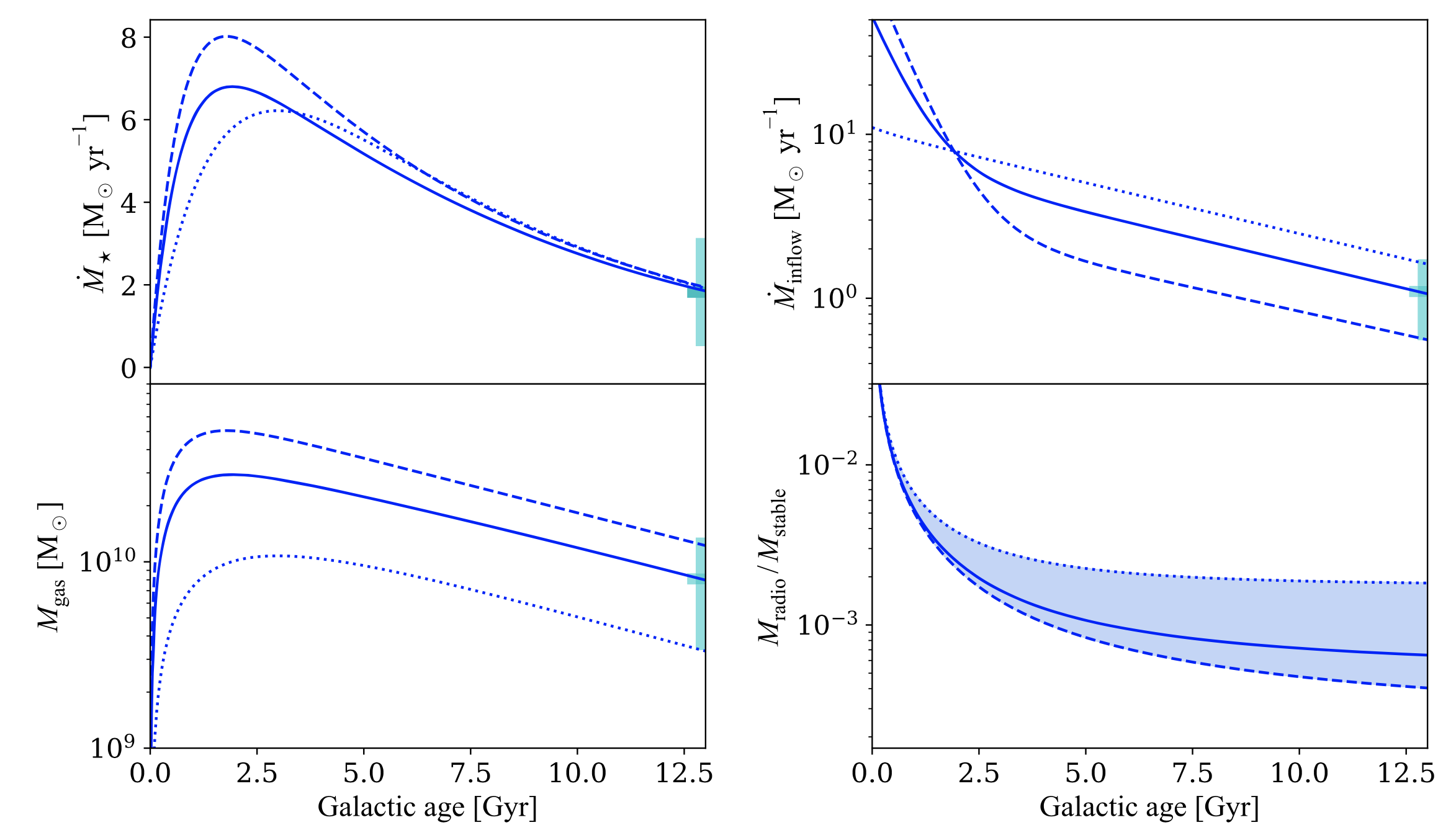}
\caption{{Previously} 
 published as Figure 6 in \cite{cote19PaperI}, \copyright AAS, reproduced with
permission. Time-evolution predictions from our OMEGA+ GCE framework for the Milky Way. Top left: {the} star formation rate $\dot{M}_*$. Top right: {the} gas inflow rate {$\dot{M}_{\rm inflow}$}. Bottom left: {the} mass of gas $M_{\rm gas}$. Bottom right: 
the abundance ratio between an SLR nucleus and a reference stable isotope {$M_{\rm radio}/M_{\rm stable}$}. The vertical bands near 13\,Gyr represent observational data taken {from \cite{kubryk15b}.} 
 In this framework, solar metallicity is reached at 8.2 Gyr. {Our best GCE model (solid line) represents the case where the current observational values of $\dot{M}_*$, $M_{\rm gas}$, and $\dot{M}_{\rm inflow}$ are all fitted. The other two lines correspond to two more GCE models we explored to cover the observational uncertainties in $M_{\rm gas}$ and $\dot{M}_{\rm inflow}$. These produce a minimum (dashed line) and a maximum (dotted line) value of $M_{\rm radio}/M_{\rm stable}$.} \label{fig:gce_a}}
\end{figure}

The first step in the development of our framework was to quantify the reliability of GCE predictions, given the uncertainties in the observational constraints that we used to calibrate our Milky Way models (see Figure~\ref{fig:gce_a}). In particular, we generated a range of numerical predictions for the abundances ratios between SLR nuclei and their reference isotopes by varying input parameters such as the temporal profile of the star formation history of the Galaxy, the total stellar mass formed, the present star-to-gas mass ratio, and the galactic inflow and outflow rates. Accounting for all of these sources of variations, we found that our predictions were uncertain by a factor of $\sim$\,3.6 at the time of the ESS (see bottom-right panel of Figure~\ref{fig:gce_a}). We also provided an analytical solution that reproduces our predicted abundances. Using constant production yields ratios for various enrichment sources, we used this framework to provide a range of isolation times for specific SLR nuclei. To do so, we compared our predicted abundances (taken at the time the Sun formed) to the ESS values and calculated how much time was needed to freely decay our predicted values down to the ESS values. For example, using 
\iso{53}Mn/\iso{55}Mn and \iso{60}Fe/\iso{56}Fe, we derived isolation times of $17-24$\ and $10-15$\,Myr, respectively. We could not derive isolation times for \iso{26}Al/\iso{27}Al because our predictions fell below the ESS value, which confirms the need for local source for this isotope { (for example, the massive star winds investigated in Figure~\ref{fig:winds})}. Furthermore, isolation times derived for \iso{92}Nb and \iso{146}Sm were inconsistent between each other, when using the assumption that both isotopes came from SNeIa ({which confirmed} the result of \cite{lugaro16pprocess}).

Our GCE code OMEGA+ assumes homogeneous mixing and a continuous stellar enrichment process. However, when studying the chemical evolution of a given parcel of gas (i.e., the parcel of gas from which the Sun would eventually form) within the Milky Way, the stellar enrichment process should be considered as made up by discrete events. A new enrichment event (e.g., a nearby supernova explosion) will bring freshly synthesised isotopes and increase the overall abundance of SLR nuclei. Following such an enrichment event, the SLR nuclei freely decay until another enrichment event occurs nearby. To account for such an evolution, we extended our GCE codes and developed a statistical Monte Carlo framework that quantifies the confidence level of our GCE predictions given the stochastic nature of local enrichment events \cite{cote19PaperII,yague21PaperIII}. 

\begin{figure}[H]
\includegraphics[width=8.5 cm]{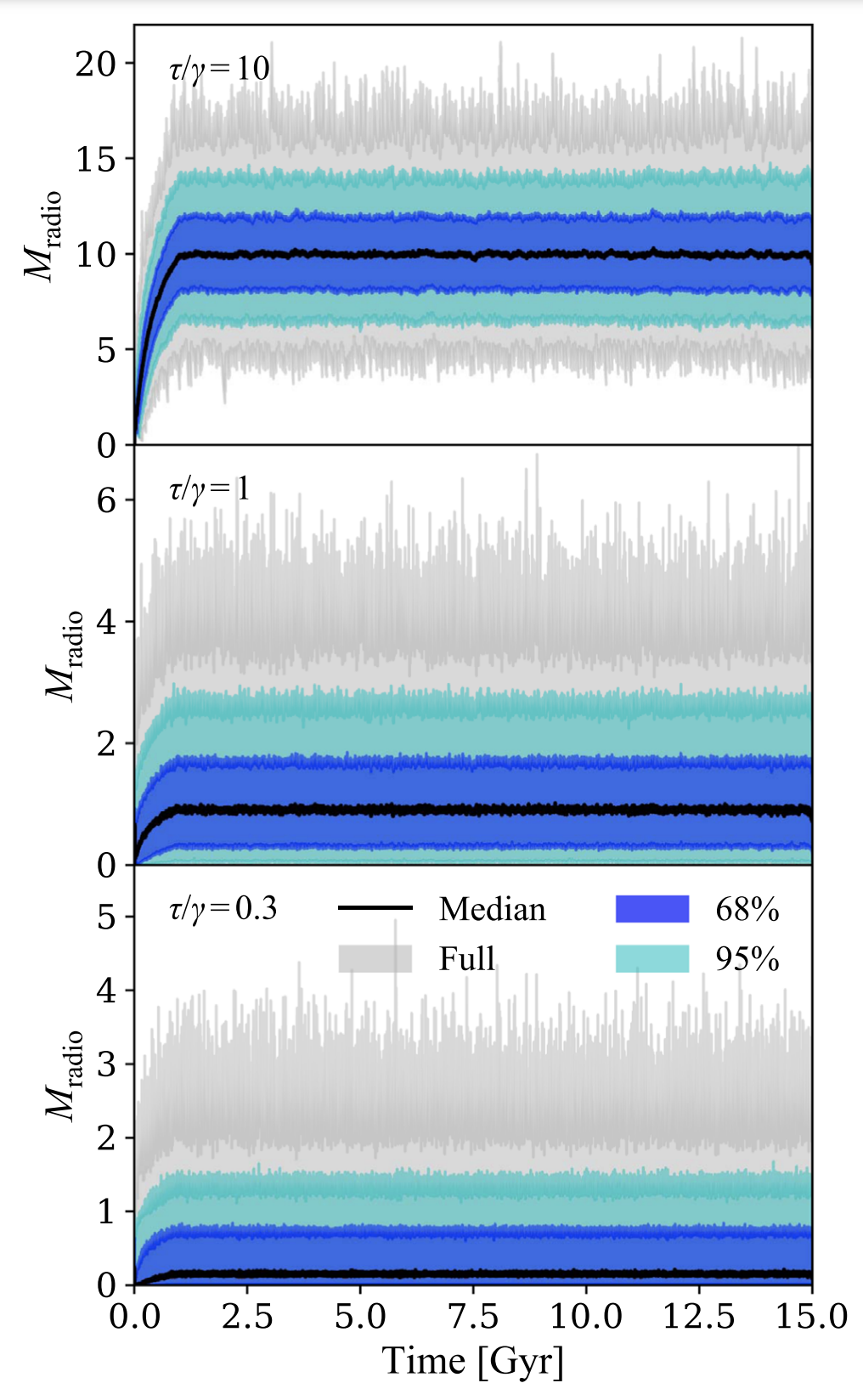}
\caption{Previously published as Figure 6 in \cite{cote19PaperII}, \copyright AAS, reproduced with
permission. Statistical distribution of {{the mass of an}} SLR {in the ISM ($M_{\rm radio}$)} as a function of galactic age, assuming each enrichment event ejects 1\,M$_\odot$ of radioactive material. Different panels show results from our stochastic Monte Carlo framework using different $\tau/\gamma$ values, representing ratios between the SLR mean-life and the average time interval between enrichment events.  \label{fig:gce_b}}
\end{figure}

We first defined a constant time interval between the formation of enrichment progenitors. We then calculated delay times, defined as the time interval between the formation of progenitors (e.g., binary neutron star systems) and their enrichment event (e.g., neutron star merger) by randomly sampling delay-time probability distribution functions. We found that the spread (or width of our confidence levels) depends on the ratio between the mean-life ($\tau$) of the considered SLR nucleus and the average time interval ($\gamma$) between enrichment events \cite{cote19PaperII}. When $\tau/\gamma > 2$, the ejecta of new enrichment events tends to pile up on the SLR abundances synthesised by previous events. In other words, SLR nuclei do not have time to completely decay before the occurrence of a new enrichment event. In this regime, the abundance of an SLR nucleus reaches a steady-state value defined by the product of $\tau/\gamma$ times the nucleosynthetic yields and has a spread that can be properly quantified with statistical confidence levels (top panel of Figure~\ref{fig:gce_b}). Overall, we found that the enrichment stochasticity adds at most an extra 60\% uncertainty to the much larger GCE uncertainties (factor of $\sim 3.6$) derived from the homogeneous framework described above. Instead, when $\tau/\gamma < 1$, SLR nuclei significantly decay before a new enrichment event occurs (bottom panel of Figure~\ref{fig:gce_b}). In this regime, extremely low values of the abundance can be reached and the spread of an SLR abundance cannot be properly quantified.

In addition to investigating the timescales at which the ESS was enriched with SLR nuclei, our GCE framework can also be used to provide insights into the physical conditions in which these SLR nuclei were produced. In particular, we found that abundances ratios between two SLR nuclei, rather than ratios between an SLR nucleus and its stable reference isotope, are sometimes more effective in probing the astrophysical sites of SLR nuclei \cite{yague21PaperIII}. This is because they are less affected by the uncertainties associated with the long-term evolution of our Galaxy, which means that the predicted ratios become more sensitive to the nuclear astrophysics inputs. Nevertheless, such an abundance ratio still varies over time due to the different timescales at which the two SRLs freely decay between enrichment events \cite{yague21PaperIII}. However, when the half-lives of the two SLR nuclei are similar, their abundance ratio does not significantly vary with time. In that regard, using our stochastic Monte Carlo framework, we found four ratios for which our predicted uncertainties are minimal and controlled by their stellar production: \iso{129}I/\iso{247}Cm, \iso{107}Pd/\iso{182}Hf, \iso{97}Tc/\iso{98}Tc, and \iso{53}Mn/\iso{97}Tc~\cite{yague21PaperIII}.

\subsection{The SLR Isotopes Produced by the $r$ process}
\label{sec:GCEr}

We studied in more detail the case of \iso{129}I/\iso{247}Cm, which involves two SLR nuclei synthesised by the {{$rapid$} neutron-capture ($r$)} process. We found that since \iso{129}I and \iso{247}Cm have almost the same half-life (15.7 and 15.6\,Myr, respectively), their abundance ratio did not change significantly since the last event that enriched the pre-solar nebula \cite{cote21science}. In addition, given the rarity of $r$-process events ($\tau/\gamma < 1$), we found that the meteoritic \iso{129}I/\iso{247}Cm ratio is likely dominated by the contribution of one event. This ratio can thus be seen as a direct window into the nucleosynthesis of the last $r$-process event that occurred $\sim$100--200\,Myr~\endnote{This time from the last $r$-process event is derived self-consistently using the two ratios \iso{129}I/\iso{127}I and \iso{247}Cm/\iso{235}U.} prior to the formation of the Sun. In collaboration with an international team of experts in $r$-process nucleosynthesis, we found that the last wave of $r$-process production likely occurred in moderately neutron-rich conditions \cite{cote21science}. The \iso{129}I/\iso{247}Cm ratio represents a unique example of how GCE uncertainties could be completely removed from the picture, following our detailed uncertainty analysis. While numerical predictions are still sensitive to the large nuclear physics and astrophysical uncertainties, the reduced layer of uncertainties brings us closer to the astrophysical origin of these SLR nuclei. 

\subsection{The GCE of SLR Isotopes Produced by AGB Stars}
\label{sec:GCEs}
\begin{figure}[H]
\includegraphics[width=13.5 cm]{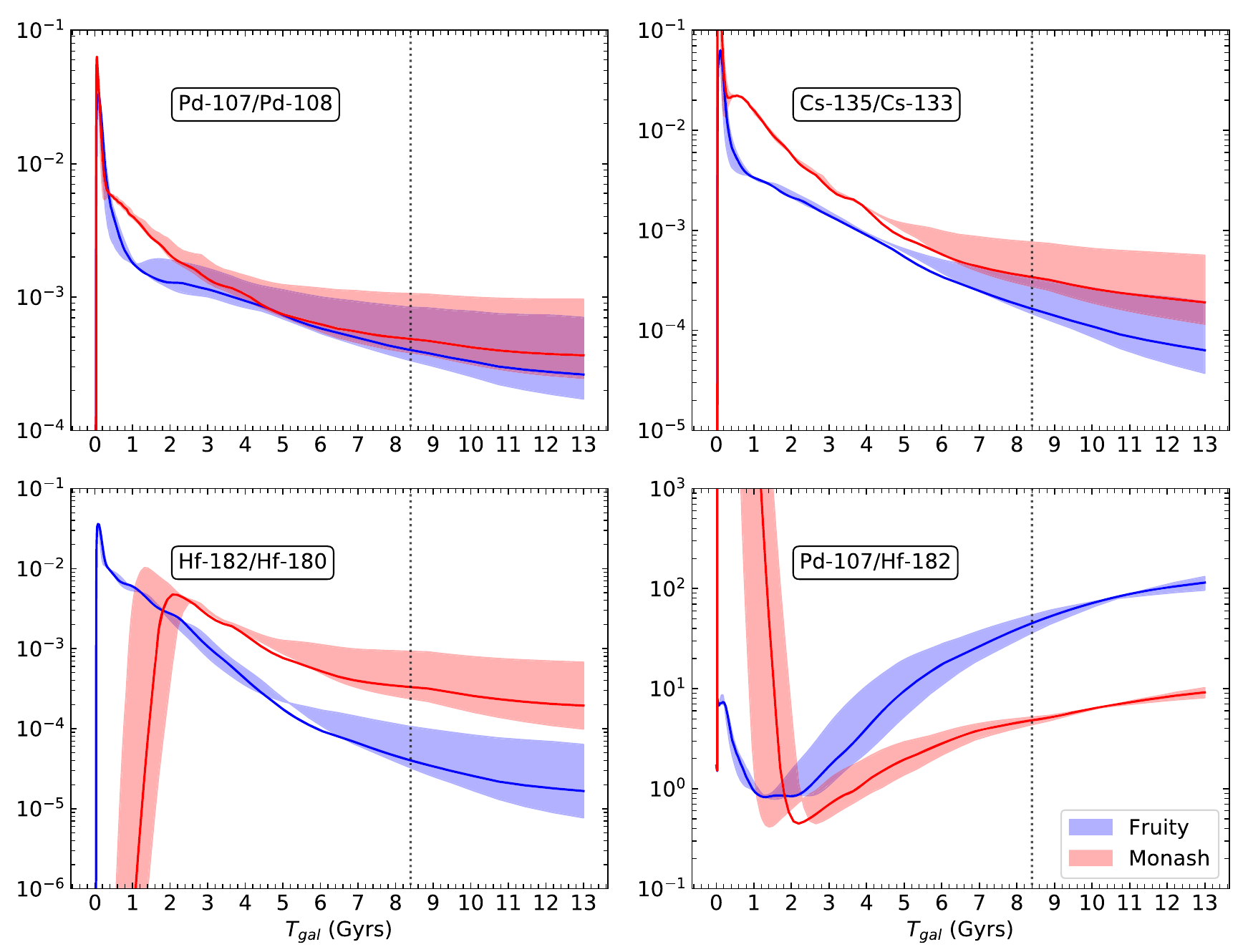}
\caption{{{Previously} 
 published as Figure 3 in \cite{trueman21}, \copyright AAS, reproduced with
permission. Time-evolution predictions from our OMEGA+ GCE framework for the Milky Way of the SLR isotopes produced by AGB stars, relative to their stable reference isotope (the two top and the bottom left panel). These panels are the realisation of the bottom right panel of Figure~\ref{fig:gce_a} for the three specific isotopes, where the solid line is the best fit GCE and the colored bands represent the minimum and maximum models, derived using the Fruity (blue) and the Monash (red) yields. The bottom right panel represents the evolution of the ratio of the two SLR isotopes \iso{107}Pd/\iso{182}Hf, for which we found the uncertainties from the stochastic Monte Carlo framework to be minimal. The dotted vertical line represents the time of the formation of the Sun and we decayed the values of the ratios at that time to their corresponding ESS values to calculate the isolation time.} \label{fig:gce_s}}
\end{figure}

{Using }{our} updated {GCE} framework, we also investigated in more detail the origin of SLR nuclei synthesised by the {{$slow$} neutron-capture ($s$)} process {in AGB stars} \cite{trueman21}. {To this aim}, we incorporated {two different sets of} mass- and metallicity-dependent AGB yields, {the Fruity \cite{cristallo09FRUITY2M,cristallo11FRUITY} and the Monash \cite{karakas16} yields}. Then, we followed the chemical evolution of the  \iso{107}Pd/\iso{108}Pd, \iso{135}Cs/\iso{133}Cs, and \iso{182}Hf/\iso{180}Hf ratios also accounting for GCE uncertainties (Figure~\ref{fig:gce_s}). 

{The main difference between the Fruity and the Monash models is the treatment of the decay rate of \iso{181}Hf, which is the branching point on the $s$-process path responsible for the production of \iso{182}Hf. The Monash models include the terrestrial value of the decay rate. This is to follow the results of Reference~\cite{lugaro14science}, who showed that the previously recommended  
faster decay of \iso{181}Hf at stellar temperatures was based on a wrong assignment \cite{bondarenko02}. The Fruity models include the previous faster decay of \iso{181}Hf and, therefore, produce a much lower \iso{182}Hf abundance than the Monash models.} 

When assuming $\tau/\gamma > 2$, and adding the Monte Carlo stochastic uncertainties to the values of the ratios shown in Figure~\ref{fig:gce_s}, {we found self-consistent isolation times, between 9 and 26\,Myr, using all three abundance ratios and the Monash yields. The \iso{107}Pd/\iso{108}Pd and \iso{135}Cs/\iso{133}Cs ratios calculated using the Fruity yields are also in agreement with this range of values.} From considering the \iso{107}Pd/\iso{182}Hf ratio, we also identified a potential missing nucleosynthesis source of Pd in our GCE code. {Recently, some of us \cite{cseh22} compared the abundances of a large sample of Ba stars, the binary companions of AGB stars, to predictions from Monash and Fruity models. In general, we found that many such Ba stars have excesses in the elements Nb, Mo, and Ru that are not explained by current AGB $s$-process models. Although it has not been possible to observe Pd yet in Ba stars, this element is located just after Ru. Therefore, it may also be present there in excess of the AGB models. This would be in qualitative agreement with the missing \iso{107}Pd relative to \iso{182}Hf abundance we have found in Trueman et al. \cite{trueman21}.}

When assuming $\tau/\gamma < 1$, instead, we could calculate the time that elapsed between the formation of the Solar System and the last $s$-process enrichment event that polluted the pre-solar nebula. In particular, we found a self-consistent elapsed time of 25.5\,Myr when assuming a 2\,M$_\odot$ AGB model with an initial composition of $Z=0.01$.






\section{Ongoing and Future Work}
\label{sec:final}

There are several current investigations ongoing and planned within the RADIOSTAR project to move towards a complete picture of the stellar origin of SLR nuclei and their presence in the ESS. We present a brief summary of each of them below, with a prospective on the future work and impact beyond the project.

\begin{figure}[H]

\begin{adjustwidth}{-\extralength}{0cm}
\centering 
\includegraphics[width=9 cm]{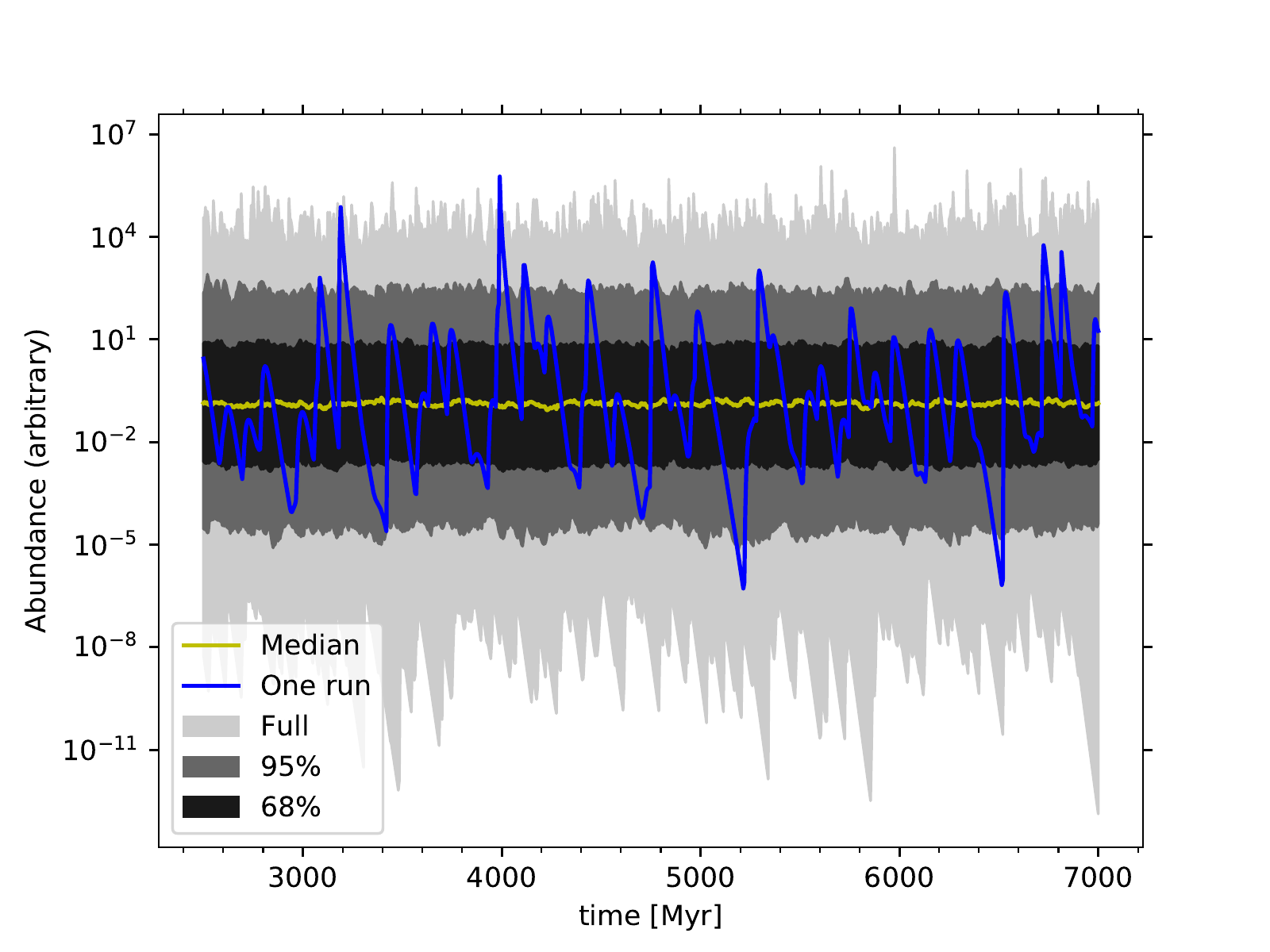}
\includegraphics[width=9 cm]{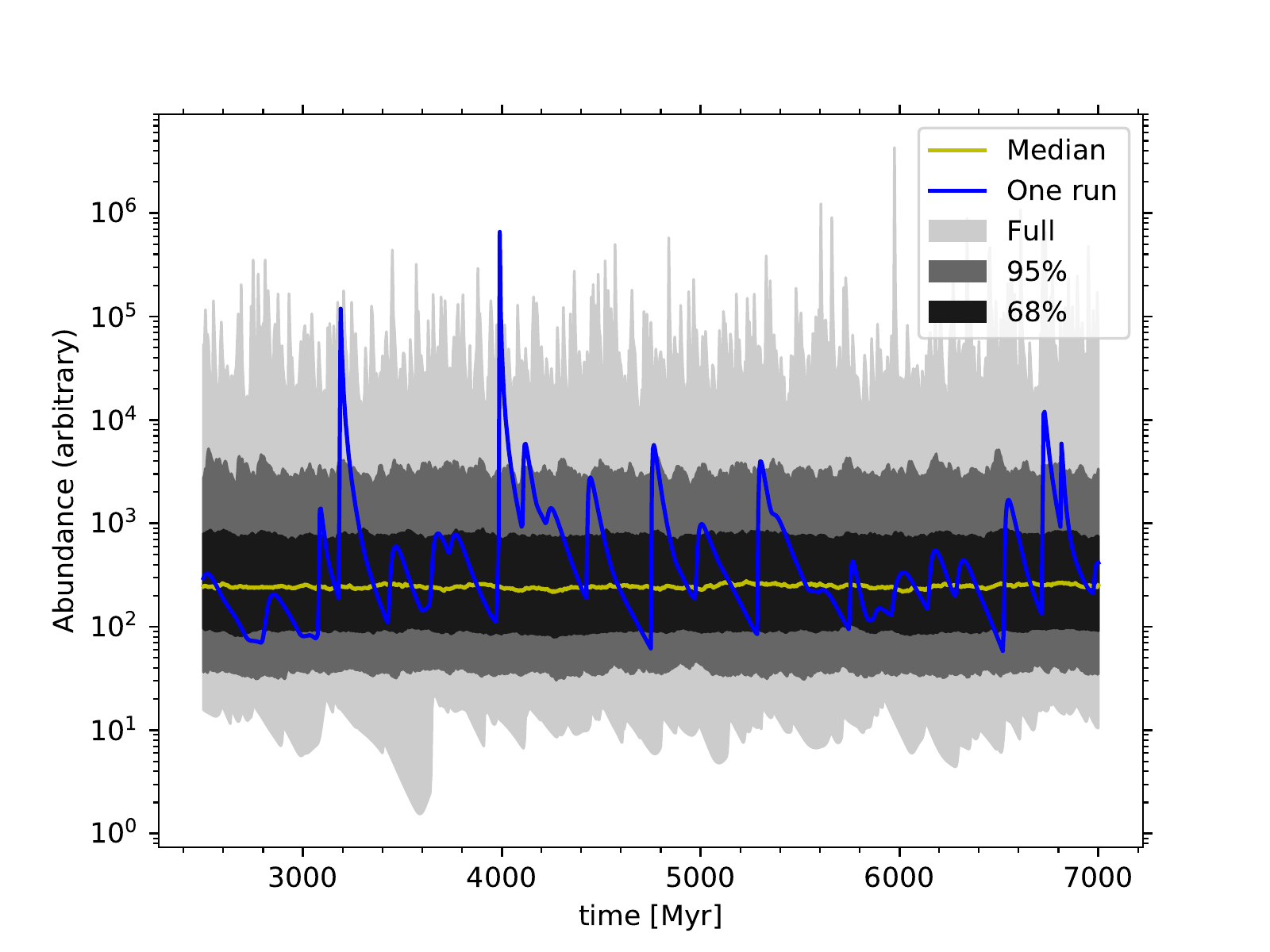}
\end{adjustwidth}

\caption{{\it Hotokezaka-style} model example calculations showing the {abundance evolution in time of an 
SLR nucleus in the ISM. For the 1000 runs calculated for this example,} the median (yellow line), 1$\sigma$ 
(black region), 2$\sigma$ (dark grey), and total (light grey) distribution are shown, as well as one 
example case (blue line). The set-up represents the case of a stellar source with a constant total galactic 
rate of 15 events per Myr (compatible to the current neutron-star merger rate), ISM mixing length 
$\alpha$=0.1, and turbulent velocity of 7 km/s. The mass of SLR nucleus ejected by each event is the 
numerical value 1~(if multiplied by \msun, then the y-axis can be considered in units of \msun). The left and
right panels represent the case of an SLR nucleus with $\tau$ = 10 and 100 Myr, respectively, and the 
abundances can be compared to each other. The median for $\tau$ = 10 Myr is roughly 3 orders of magnitude 
lower than that for $\tau$=100 Myr, because the longer $\tau$ allows for the SLR nucleus to accumulate. The 
spread, e.g., at 68\%, is 2 to 3 orders of magnitudes larger for $\tau$ = 10 Myr than for \mbox{$\tau$ = 
100 Myr}, which reflects the impact of the faster decay. The peaks in the single case, blue lines are more 
spaced out in the \mbox{$\tau$ = 100 Myr} case because the slower decay and memory build up produces a 
higher median abundance; therefore, some of the events do not have a visible impact. 
\label{fig:hotokezaka}}

\end{figure}

\begin{enumerate}[label=,labelsep=0mm]
\item {\textbf{CCSN nucleosynthesis}} 
 We are currently investigating (Lawson et al., in preparation) if our CCSN models produce a self-consistent solution for some of the SLR nuclei present in the ESS. {We are also considering} the effect on SLR abundances of further processes and uncertainties related to explosion and mixing in CCSNe beyond those already mentioned in Section~\ref{sec:stars}. Specifically, we are looking at the effects of possible ingestions of protons in the He shell \cite{pignatari2015} and of merging of different shells \cite{ritter18} just prior to the explosion. Updated investigations will still be needed on the impact of nuclear reaction rate uncertainties on the production of each SLR nucleus in CCSNe, even for the best studied $^{26}$Al and $^{60}$Fe. For instance,  {new experimental constraints have allowed us to significantly reduce the impact of the {uncertainty} of the $^{59}$Fe(n,$\gamma$)$^{60}$Fe cross section \cite{yan21fe59}, one of the main nuclear inputs for the production of $^{60}$Fe \cite{jones:19}.}

\item \textbf{Massive star winds and the production of \iso{107}Pd} We are currently calculating the production and ejection of \iso{36}Cl and \iso{41}Ca in binary systems. At the same time, we are also extending our current nuclear network to include the production of SLR nuclei up to \iso{107}Pd (Brinkman et al., in preparation). {Within} the possible scenario where \iso{26}Al, \iso{36}Cl, and \iso{41}Ca in the ESS originated from massive star winds {(Figure~\ref{fig:winds})}, \iso{107}Pd is the only heavier isotope that can also be significantly produced. As discussed in Section~\ref{sec:GCEs}, our current modelling of the $s$-process contribution from AGB stars to the galactic background can explain the currently recommended ESS value of the \iso{107}Pd/\iso{108}Pd ratio self-consistently with that of \iso{182}Hf/\iso{180}Hf. Therefore, a second contribution to \iso{107}Pd from massive star winds may result in overproduction relative to the ESS value and create a problem for this scenario. For the first time, the RADIOSTAR project can investigate together all the potential different components of \iso{107}Pd in the ESS. Within the topic of massive star winds, we also need to investigate if the produced SLR nuclei can be incorporated into dust. Dust is necessary to penetrate the ESS material and carry and deposit the SLR abundances within it. It is observed to form in the carbon-rich winds of massive binary star systems \cite{lau20}. {Finally, we note that, aside from massive star binaries and Wolf--Rayet stars, the winds of stars with mass above 100 \msun\
can also contribute significantly to the \iso{26}Al enrichment of the ISM. These stars are nearly homogeneous and can convert almost all \iso{25}Mg initially present into \iso{26}Al via proton captures~\endnote{For example, the 500\,M$_{\odot}$ model by \cite{yusof13} produces 100 more $^{26}$Al than their 60\,M$_{\odot}$ star.}. These stars are extremely rare, however, and even a few events may have a strong impact, which should be analysed in relation to the ESS.}

\item \textbf{Origin of \iso{244}Pu} Among the $r$-process SLR nuclei, \iso{244}Pu has also been observed to be present in the ESS, although its abundance is still relatively uncertain. New laboratory measurements {within our project are} aimed at better defining its ESS value and distribution (Pet\H{o} et al., in preparation). We are also considering if the same models that can explain \iso{129}I and \iso{247}Cm can also fit \iso{244}Pu (Lugaro et al., in preparation). The further complication of this isotope is that its half-life (of 80 Myr) is much longer than that of the other two isotopes; therefore, it is more likely that the abundance of this SLR nucleus carried the memory of several events in the galactic background. 

\item \textbf{Heterogeneous GCE modelling} One of the main open problems to achieve an accurate description of the abundances of SLR nuclei both within GCE and in molecular clouds, is the transport in the ISM. So far, our GCE models have been simplistic in this respect because they do not included transport. By considering stellar sources that produce both the SLR and the stable reference isotope, we have exploited the fact that the dilution factor due to transport from the source to each parcel of ISM gas must be the same for both isotopes. However, reality is more complex because such dilution factor would effectively give a different {weight} to different sources, depending on their distance and on how far the isotopes can travel. Effectively, distance and speed control the numbers of sources that contribute material to a given parcel of gas and, therefore, when considered per unit time, the parameter $\gamma$. Furthermore, the stable isotope abundance completely loses the memory of each single production event, as {material becomes well mixed} in the Galaxy within its rotation period of 100 Myr. {Instead,} the SLR nuclei keep the memory of the events that occurred locally in time and space. For these reasons, we are now developing more complex models. {We have introduced} SLR nuclei into the Inhomogeneous Chemical Evolution (ICE) code~\cite{wehmeyer15}, where mixing in the ISM is treated in three dimensions and driven by supernova explosions (Wehmeyer et al., in preparation). {We have also developed} a mixing code based on the scheme of Hotokezaka et al. \cite{hotokezaka15} where material is transported by diffusion (Yag\"ue Lop\'ez et al., in preparation). A couple of preliminary test examples of our {\it Hotokezaka-stlye} simulations are shown in Figure~\ref{fig:hotokezaka}.

\item \textbf{Chemodynamical SPH simulations} We are considering a higher level of complexity by introducing SLR nuclei also within sophisticated models of the Galaxy based on cosmological constraints \cite{kobayashi11chemodyn}. While these models cannot zoom into each single stellar source, they account for all the dynamical features of the Milky Way and provide us a more accurate description of the distribution of SLR nuclei in the Galaxy. We have introduced a number of SLR nuclei within such models and are currently running high-resolution simulations to be compared to global galactic observables, such as the \iso{26}Al $\gamma$-ray emission line (Wehmeyer et al., in preparation). 

\item \textbf{\iso{53}Mn and the contribution of Type Ia supernovae} \iso{53}Mn is a particularly interesting well-known SLR nucleus in the ESS whose abundance still needs to be analysed in terms of stellar sources and GCE evolution. The element Mn is produced most significantly in the Galaxy by SNeIa, and particularly those that reach the Chandrasekhar mass \cite{seitenzahl13,kobayashi20MnNi}, but also partly by CCSNe, also depending on the models considered. A full GCE model is required to follow the production of \iso{53}Mn and \iso{55}Mn, together with that of \iso{56}Fe, also significantly produced by SNeIa, and \iso{60}Fe, which is only produced by CCSNe instead. Such a model will help us to verify if the origin of both \iso{53}Mn and \iso{60}Fe in the ESS can be attributed to the galactic background, similarly to the longer lived $s$- and $r$-process SLR nuclei discussed in Section~\ref{sec:galaxy}.

\item \textbf{The $p$-process SLR nuclei} Finally, \iso{92}Nb and \iso{146}Sm are $p$-process isotopes for which the ESS values are well known, while their stellar origin is still unclear. The origin of $p$-process nuclei in general is still strongly debated, with many potential sources related to various types of supernova explosions. Detailed GCE models of the evolution of the abundances of these two isotopes are needed and may help in understanding the origin of the $p$ process by testing different combinations of stellar yields. Furthermore, the other two SLR nuclei synthesised by the $p$ process, \iso{97}Tc and \iso{98}Tc, have very close half-lives (4.21 and 4.2\,Myr, respectively) and could be the subject of a study similar to that which we performed for \iso{129}I/\iso{247}Cm, once more precise meteoritic abundance determinations are available for these isotopes~\cite{yague21PaperIII}.
\end{enumerate}
\vspace{-12pt}


On top of the efforts related to the modelling of the stellar sources and galactic evolution listed above, improvements in the nuclear and meteoritic experimental data would strongly help us to constrain the problem of the origin of SLR nuclei in the ESS. 
In terms of nuclear physics inputs, several advances have been made, also with the contribution of the RADIOSTAR project (see \cite{li21cs135,yan21fe59} and Laird et al. 2022, submitted to J.Phys.G.). Main inputs are still required in terms of charged-particle reactions, for example, on the crucial \iso{22}Ne($\alpha$,n)\iso{25}Mg reaction that produces the neutrons needed to make \iso{182}Hf in AGB stars.  Neutron-capture reactions are also to be improved, for example, those on \iso{36}Cl and \iso{41}Ca that control the abundance of these isotopes in massive star winds \cite{brinkman21}, as well as decay rates with their temperature and density dependence. A crucial case is that of \iso{146}Sm, whose terrestrial decay rate is still debated between the two values of 68 \cite{kinoshita12} and of 103 Myr~\cite{marks14}. {Furthermore, model predictions for} \iso{205}Pb (and, to a lesser extent, for \iso{41}Ca) also present large uncertainties {due to the almost unknown}  temperature and density dependence of its electron capture rate.

Finally, more accurate determination of the SLR ESS abundances 
would truly allow us to discard and/or favour different scenarios.
{Some crucial isotopes discussed above and to be improved} are \iso{36}Cl, \iso{41}Ca, \iso{107}Pd, and \iso{244}Pu. More resonance ionisation mass spectrometry (RIMS) data on \iso{60}Fe are also needed. {The} potentially correlated presence of different SLR nuclei (e.g., \cite{holst13}) and even with stable isotope anomalies in the same materials {would also be very useful constraints. Furthermore,} ESS values, rather than upper limits, {are still needed} for \iso{205}Pb (main $s$ origin and minor massive star wind production \cite{arnould06}), \iso{135}Cs ($r$ and $s$ origin), \iso{97}Tc ($p$ origin), and \iso{98}Tc (main $p$ origin and minor massive star wind production~\cite{arnould06}). {These would allow us to confirm} the origin of the $s$-process SLR nuclei and to link the $p$ process to Chandrasekhar mass SNeIa, which also produce \iso{53}Mn, whose half-life is very similar to that of \iso{97}Tc and \iso{98}Tc \citep{yague21PaperIII}. Finally, \iso{126}Sn is a potentially interesting very short-lived (0.23 Myr) isotope of main $r$ origin and minor CCSN production, for which we only have an ESS upper limit. For all these isotopes, it is outstandingly difficult to derive accurate and precise ESS values for several reasons, from their very low abundance (e.g., the Tc isotopes) to their volatility (for Cs and Pb) to problems with age determinations of the meteorite (for \iso{107}Pd). Any analytical improvement will help us to better exploit our analysis of their astrophysical origin.


\authorcontributions{All the authors have contributed to the conceptualisation, methodology, software, 
validation, formal analysis, and investigation. The original draft was prepared by M.L, {B.C. (Benoit C\^ot\'e), and M.P. (Marco Pignatari),} 
 with Figure~\ref{fig:gce_b} contributed by A.Y.L. All the authors contributed to the review and editing of the paper. M.L. contributed to the supervision, project administration, and funding acquisition for the project. All authors have read and agreed to the published version of the manuscript.}

\funding{This research was funded by ERC via CoG-2016 RADIOSTAR (Grant Agreement 724560).}

\institutionalreview{Not applicable.} 


\informedconsent{Not applicable.} 


\dataavailability{{Not applicable.} 
} 

\acknowledgments{We would like to thank Krisztina Botos for financial management of the project and Evelin B\'anyai for IT management and software development. We also thank all the colleagues who have visited our team since September 2017, in order of time: Ertao Li, Ulrich Ott, Waheed Akram, Martin Bizzarro, Georges Meynet, Roland Diehl, Reto Trappitsch, Melanie Hampel, Yusuke Fujimoto, Xiaodong Tang, Kuoang Li, Kai Zuber, Etienne Kaiser, Marius Eichler, Jenny Feige, Alessandro Airo, Stein Jacobsen, Akshat Garg, Claudia Travaglio, Moritz Pleintinger, Nicole Vassh, and Andre Sieverding. We are grateful for the time they have spent with us and all their help and discussion. We are also grateful for discussions with Maria Schoenbachler and Mattias Ek. Finally, we thank Sara Palmerini for the opportunity of presenting this paper and Maurizio Busso for his life-long inspiration on the study of the topic of radioactive nuclei.}

\conflictsofinterest{The authors declare no conflict of interest. The funders had no role in the design of the study; in the collection, analyses, or interpretation of data; in the writing of the manuscript; or in the decision to publish the~results.} 

\abbreviations{Abbreviations}{The following abbreviations are used in this manuscript:\\

\noindent 
\begin{tabular}{@{}ll}
AGB star 
 & asymptotic giant branch star \\
CCSN (CCSNe) & core-collapse supernova (core-collapse supernovae) \\
ESS & early Solar System\\
GCE & galactic chemical evolution\\
ISM & interstellar medium \\
$rapid$ neutron-capture process & $r$ process \\
$slow$ neutron-capture process & $s$ process \\
\end{tabular}}

\noindent 
\begin{tabular}{@{}ll}

SLR & short-lived radioactive \\
SNIa (SNeIa) & supernova Type Ia (supernovae Type Ia) 
\end{tabular}

\begin{adjustwidth}{-\extralength}{0cm}
\printendnotes[custom]
\end{adjustwidth}

\begin{adjustwidth}{-\extralength}{0cm}


\reftitle{References}



\end{adjustwidth}
\end{document}